\documentclass[12pt]{iopart}  
\usepackage{listings}
\usepackage{hyperref}
\usepackage{color}
\usepackage{soul}
\usepackage{graphicx}

\def\EQ#1{\begin{eqnarray}#1\end{eqnarray}}

\begin{document}

\title[]{To quantum or not to quantum:\\ 
towards algorithm selection in near-term quantum optimization}

\author{Charles Moussa$^1$, Henri Calandra$^2$, and Vedran Dunjko$^1$}

\address{$^1$ LIACS, Leiden University, Niels Bohrweg 1, 2333 CA Leiden, Netherlands}
\address{$^2$ TOTAL SA, Courbevoie, France}

\eads{\mailto{c.moussa@liacs.leidenuniv.nl}, \mailto{henri.calandra@total.com}, \mailto{v.dunjko@liacs.leidenuniv.nl}}

\begin{abstract}
The Quantum Approximate Optimization Algorithm (QAOA) constitutes one of the often mentioned candidates expected to yield a quantum boost in the era of near-term quantum computing.  
In practice, quantum optimization will have to compete with cheaper classical heuristic methods, which have the advantage of decades of empirical domain-specific enhancements. 
Consequently, to achieve optimal performance we will face the issue of algorithm selection, well-studied in practical computing. Here we  introduce this problem to the quantum optimization domain.

Specifically, we study the problem of detecting those problem instances of where QAOA  is most likely to yield an advantage over a conventional algorithm. As our case study, we compare QAOA against the well-understood approximation algorithm of Goemans and Williamson (GW) on the Max-Cut problem. As exactly predicting the performance of algorithms can be intractable,  
we utilize machine learning to identify when to resort to the quantum algorithm. We achieve cross-validated accuracy well over 96\%,
 which would yield a substantial practical advantage.  
In the process, we highlight a number of features of instances rendering them better suited for QAOA. While we work with simulated idealised algorithms, the flexibility of ML methods we employed provides confidence that our methods will be equally applicable to broader classes of classical heuristics, and to QAOA running on real-world noisy devices.
\end{abstract}

\vspace{2pc}
\noindent{\it Keywords}: Algorithm Selection, Combinatorial Optimization, Quantum Algorithms

\section{Introduction}
\label{intro}

Quantum computing in the near-to-mid term -- the so called NISQ era  \cite{Preskill2018quantumcomputingin} -- is likely to be  limited in many ways, including size (qubit numbers), gate fidelities, architecture (qubit connectivity) and qubit life-times (coherence times).
Consequently, much effort is dedicated to the development of algorithms that can work around some of these constraints. 
Approaches based on parameterized quantum circuits -- such as variational quantum eigensolvers \cite{VQE}, the quantum approximate optimization algorithm (QAOA) \cite{QAOA}, and some flavours of quantum machine learning \cite{PQC,Lamata2020}, stand out as some of the most likely NISQ-suitable algorithm families. They are suitable in part because the quantum computational depth, dictating required coherence times, can be treated as a tunable ``hyper-parameter'', and made to fit the device constraints. In this work, we specifically focus on the QAOA algorithm, which consists of a quantum circuit  of a user-specified depth $p$, inducing  $2p$ \emph{QAOA parameters} which are tweaked to solve a combinatorial optimization problem. QAOA is interesting from the perspectives of approximation, and heuristic optimization.

\noindent \textbf{Approximation with QAOA -} Initially, the QAOA algorithm was designed as an \emph{approximation algorithm}, i.e., an algorithm intended to provide an efficient, but approximate solution to an otherwise intractable problem. Approximation algorithms differ from \emph{heuristic} methods in that they provide mathematically provable a-priori guarantees on the quality of the solution, whereas heuristics utilize expert knowledge, and domain-specific operators to achieve good performance in practice, with less concern regarding theoretical worst case bounds.
For instance, one of the most natural applications of the QAOA algorithm is to solve so called MaxCut problems, where the task is to identify a bipartition of a given graph, which maximizes the number of edges (the \textit{cut}) crossing the two sets. Finding the actual maximum of the cut $C_{max}$ is NP-hard \footnote{Technically, the decision variant of the problem, deciding whether the cut is larger than some integer $k$, is NP-complete. }, but the problem allows non-trivial polynomial-time approximation. The best known classical approximation algorithm, of Goemans and Williamson (GW) \cite{Goemans:1995:IAA:227683.227684}, guarantees an \emph{approximation ratio} of $\alpha \approx 0.878$ (referred to as the GW bound). In the case of randomized algorithms like GW, this means that the value of the output cut is at least $\alpha C_{max}$ with high probability.
Under which conditions better efficient approximations can be achieved is a hard open question, and depends on details. For instance, it is known that achieving an approximation ratio above $16/17\approx 0.94$ for general graphs is NP-hard \cite{Hastad2001} (so as hard as finding the exact solution), and it is possible that the actual NP-hard bound may be $\alpha$. 
For more special graphs the GW can do better, e.g, for 3-regular graphs, the ratio of $0.932$ is achieved. 
To use QAOA for optimization, a critical step is finding the optimal classical circuit parameters for a given depth. For a constant depth, this can be done in polynomial time, yielding  a poly-time approximation algorithm \footnote{Without constant depth limits, it is known that QAOA can achieve exact solutions, but the classical parameter optimization then can take exponential time.}.
The question of whether QAOA can improve over GW for any constant depth was open for a number of years, and has been resolved very recently, in the negative \cite{variationalObstacles}.
However, already at depth $p=1$ it achieves the ratio of 0.6924. While falling short of the GW bound, this is a highly non-trivial result, as prior to GW, the best algorithms achieved only $1/2 + o(1)$\footnote{In other words, the achieved approximation ratio is below $1/2 + \epsilon$ for any constant $\epsilon>0$.}, even after decades of efforts. This makes QAOA an interesting algorithm, and allows the possibility that QAOA performs better than GW (or some other classical algorithm) on some classes of instances.
Setting the theoretical approximation aspects aside, this suggests the substantial practical potential of QAOA as a \emph{heuristic optimizer}.

\noindent \textbf{Heuristic optimization with QAOA -} The above-mentioned theoretical results motivate studying how to best use QAOA for practical optimization problems via numerical (and real-device) experiments. As mentioned previously, to actually use QAOA one needs to find the optimal classical parameters. Although finding exact optima for any constant depth can be done in polynomial time \cite{QAOA}, the actual overheads are prohibitive, requiring billions of evaluations already for small graphs and depth in low single digits.
Almost as much research has been dedicated to developing heuristics and analytical techniques to find good circuit parameter values for QAOA \cite{qaoa_perf,qaoaConcentrates,multistartQAOA,learninglearn}, as has been to actual evaluation of the performance of QAOA \cite{Crooks2018b}. This is justified as any evaluation of QAOA will be in part limited by the limitations of the classical optimizer. The early studies are promising, showing, e.g., that QAOA with depth $p=8$  significantly outperforms GW on a set of Erd\H{o}s-R\'{e}nyi random graphs (with edge probability 0.5) \cite{Crooks2018b}. A somewhat related theoretical study identifies that there exist instances where QAOA will outperform quantum and simulated annealing methods \cite{qaoavsannealing}.  

Perhaps the first main issue with QAOA is that many open questions still remain, e.g., regarding the comparison of QAOA with other heuristic optimization methods, on other graphs, with different optimizers, and with varying levels of experimental (or simulated) noise -- in part as simulating, or actually running it on a real device QAOA is computationally costly, for larger graphs, preventing large-scale analyses. We note that the issue of noise was addressed e.g. in \cite{Guerreschi2019,analysisNoise1,analysisNoise2}.

The second issue is more subtle and pertains to the importance of predicting when to use QAOA, on an instance-to-instance basis. This capacity would allow us to build better hybrid solvers, with possible implications in combined approximation algorithms (see e.g. \cite{Goemans2009CombiningAA}) or in the heuristic setting (see e.g.  \cite{aaAndGA}). Note that despite these early successes of QAOA on, realistically speaking, very small instances, QAOA cannot beat even the bare GW algorithm on all instances at any constant depth \cite{hastings,variationalObstacles}. This holds even when assuming ideal parameters and zero noise. We note that this result was not unexpected, as previously known results imply that, under the so-called unique games conjecture (UGC \cite{Khot:2002:PUG:509907.510017,Khot:2007:OIR:1328722.1328735}), QAOA could not beat GW unless BQP contains NP.

QAOA should thus not be expected to universally win even against GW, let alone all standard heuristic methods, optimized for certain classes of problems. Further, given the expected substantially higher expenses of running actual quantum hardware, in practice we will need to leverage this additional cost against the actual improvement (potentially) offered by QAOA over less costly classical algorithms. 
Thus one should tackle the challenge of \textit{algorithm selection} in quantum optimization, which in the classical realm is mostly concerned with using the fastest algorithm for a given instance \cite{Kotthoff2016}. As we clarify presently, detecting which instances are ``quantum-suitable'' is likely itself a hard problem, requiring a heuristic treatment. Investigating whether such a detection is possible using machine learning techniques is the key theme of this work.

Here, we investigate such a detection method for the case of the QAOA  versus the GW algorithm. A thoughtful critic may point out that a) since the overall approach is heuristic, and b) since of QAOA is most likely going impact the heuristic domain, detection-of-advantage strategies against classical heuristic algorithms would be more relevant. While we absolutely agree with this sentiment, there are important reasons we opted nonetheless to focus this first investigation of its type on the QAOA/GW setting.
The main reason is pragmatic:  at present QAOA can only be run on  small instances -- as defined in in classical literature, those with  fewer than 100 nodes in \cite{DunningEtAl2015}. In our case, we are indeed limited at 24 nodes.  Mainstream classical heuristics can quite easily attain optima in these cases, which prevents any reasonable analysis. Comparing against purposefully bad algorithmic choices would have been equally uninformative.  In contrast, GW algorithm provides the right combination solid, yet imperfect performance on reachable instance sizes for our purposes (despite the fact that it has strong provable worst-case performance bounds rendering it, technically, an approximation algorithm).
As a potential secondary advantage, the ML techniques employed may provide hints for further theoretical insights in the performances of these two  well-studied algorithms.

We highlight that since our actual object of study is the capacity of  ML methods for algorithm selection (where one algorithm is QAOA), the choice of the second algorithm is not central. Further, thanks to the nature of ML approaches, our technique is likely flexible enough to be extended to settings with genuine heuristic algorithms, as soon as instance sizes allow for meaningful training sets.

\noindent\textbf{Contributions -} In this work we contribute in these directions, as follows:
\begin{itemize}
\item we compare the performance of QAOA with GW on a different family of graphs, namely 4-regular graphs;
\item we specify two pragmatically motivated criteria specifying when QAOA should be used over GW;
\item we design a machine learning model which achieves over 96\% and over 82\% balanced accuracy in predicting which MaxCut instances satisfy the above criteria, respectively, which can be used for algorithm selection;
\item we highlight a number of graph properties which are strongly correlated with a QAOA advantage, which may help guide further theoretical analyses.
\end{itemize}

\par 
The structure of the paper is as follows. Section \ref{background} provides the necessary background on GW and QAOA.  In section \ref{QAOAperfs}, we detail the simulation performed, comparing QAOA and GW, and which we used to fix the criteria when one algorithm should be chosen over another, and to generate the datasets needed for machine-learning-based algorithm selection. In section \ref{AlgoSel}, we describe the algorithm selection model, and discuss the critical features (characteristics) of MaxCut instances which influence which algorithm does better. We conclude our paper with a discussion in section \ref{discuss}.

\section{Background}
\label{background}
The MaxCut problem is one of the famous 21 NP-complete problems identified by Karp, and it naturally occurs in computer science and physics. Consider a graph $G$ over a vertex set $V$, and edge set $E$, with $N$ vertices. The problem is to find a subset $S \subset V$ such that the number of edges between $S$ and $V\setminus S$ is maximized. The set of edges between $S$ and $V \setminus S$ is called a \emph{cut}, and we will denote the size of the maximal cut with $C_{max}$. The problem naturally extends to weighted graphs, where the objective is to maximize the total weight of the edges of the cut.
A common phrasing of the MaxCut problem is as follows. Let $w_{ij}$ be the weight associated to the edge $(i,j) \in E$ (1 in the case of an unweighted graph), then the MaxCut problem is to identify the bitstring $\mathbf{z} = (z_1,\ldots, z_N) \in \{-1,1 \}^N$ where $ z_i = 1 \Leftrightarrow i \in S $ maximizing
\EQ{
C(\mathbf{z}) =  \sum_{( i,j) \in E} w_{ij} (1- z_i z_j) /2. \label{costc}
}
This quadratic binary optimization (QUBO) formulation connects the MaxCut problem with the task of finding ground states of classical spin Hamiltonians.
As phrased, the MaxCut problem is to identify the cut set ($S(z) = \{i | z_i = 1 \}$ s.t. $C(z) = C_{max}$), but, even the problem of identifying the cut size (distinguishing the $max$ versus the $argmax$ problem for $C(\textbf{z})$) is computationally difficult.
The MaxCut problem is in the APX class \cite{PAPADIMITRIOU1991425}, the set of NP optimization problems that allow polynomial-time approximation algorithms with approximation ratio bounded by a constant).

\subsection{Goemans-Williamson Algorithm for MaxCut}

The best classical approximation algorithm for MaxCut is that of  Goemans and Williamson \cite{Goemans:1995:IAA:227683.227684} (GW). It is based on solving a related relaxation of the problem -- intuitively, we solve an optimization problem in the continuous domain  $[-1,1]^N$ as a semi-definite program, giving a real-valued solution with the associated cost $C_{rlx}$. Then a sampling process which  depends on the probabilities specified by the entries in the real-valued solution  (a \emph{random projection routine}) generates a feasible bitstring solution.
Note that the cost of the found bitstring is upper bounded by $C_{rlx}$, and in fact it can happen that $C_{max} < C_{rlx}$.
 
Nonetheless, the expected value attained by the cost function using this procedure is provably lower bounded by $\alpha C_{max}$ with $\alpha \approx 0.878$, for all input graphs - in other words, it is an $\alpha-$approximation algorithm.

While better approximation bounds are possible for special graphs (e.g. for 3-regular graphs \cite{Hastad2001}, or for graphs with large cuts \cite{Goemans:1995:IAA:227683.227684}), assuming the UGC \cite{Khot:2002:PUG:509907.510017}, GW provides the best possible polynomial-time approximation, unless P=NP.

\subsection{Quantum Approximate Optimization Algorithm for MaxCut}
\label{qaoa}

The QAOA approach is inspired by adiabatic quantum computing, and as the first step, the classical cost function is encoded in a quantum Hamiltonian defined on $N$ qubits by replacing each variable $z_i$ in equation \ref{costc} by the single-qubit operator $\sigma_i^z$, and omitting the constant shift:
\EQ{ H_C = \sum_{(i,j) \in E} \frac{w_{ij}}{2} \sigma_i^z \sigma_j^z .} Note that this corresponds to a diagonal operator w.r.t. the computational basis, with $\langle \mathbf{z} \vert H_C|\mathbf{z}\rangle = w -  C(\mathbf{z})$, where $w= \frac{1}{2} \sum_{(i,j)\in E} w_{ij}$ is half of the total weight of the graph.
The second ingredient is the so-called \emph{mixer Hamiltonian} $  H_B = \sum_{j=1}^N  \sigma_j^x$, corresponding to the typical initial Hamiltonian in an adiabatic quantum computing protocol, whereas $H_C$ corresponds to the target Hamiltonian, and the bitstring corresponding to the ground state of $H_C$ also maximizes $C(\mathbf{z})$.

Instead of resorting to an adiabatic method, in QAOA we implement the (time-independent) time-evolution interchangeably using $H_B$ and $H_C$, yielding a quantum circuit of depth $O(p)$ for constant-degree graphs, so critically independent of the graph size 
(see e.g. \cite{hadfieldthesis}). This circuit is then applied to a fixed initial state, generating:
\EQ{  |\mathbf{\gamma},\mathbf{\beta}\rangle = e^{-i\beta_p H_B} e^{-i\gamma_p H_C} \cdots e^{-i\beta_1 H_B} e^{-i\gamma_1 H_C} |+\rangle^{\otimes N},  }
defined by the $2p$  parameters $\gamma_i,\beta_i, i=1...p$ (corresponding to the number of times of evolution of the individual Hamiltonians).
We will refer to these parameters also as \emph{the QAOA angles} as they map onto the angles in parameterized gates of the corresponding quantum circuit. 

Such a quantum state when measured yields a probability distribution over all possible bitstrings \footnote{In the limit of infinite depth, the distribution will converge to a distribution with full support in the optima.}. The classical optimization challenge of QAOA is to identify the sequence of parameters $\mathbf{\gamma}$ and $\mathbf{\beta}$ so as to maximize the expected value of the cost function from the measurement outcome, i.e. the quantity $ r= F_p(\mathbf{\gamma},\mathbf{\beta}) = w - \langle \gamma,\beta| H_C | \gamma,\beta \rangle$.

If we denote the optimal parameters maximizing $F_p$ with $ \mathbf{\gamma}^{\ast},\mathbf{\beta}^{\ast}, $ the approximation ratio of QAOA \emph{achieved on the given instance} is given by:
\begin{equation}
r^\ast= \frac{F_p( \gamma^{\ast},\beta^{\ast})}{C_{max}}.
\end{equation}

For this value to be computable, $C_{max}$ for the given graph must be known. Note that it is known that if $p \to\infty$, then we have $r^\ast \to 1$ \cite{QAOA}.
This quantity -- the expected achieved approximation ratio -- is equally defined for the GW and for the QAOA algorithm, and we use it as our central figure of merit regarding the performance of the two algorithms on MaxCut instances. In the next section, we analyze the performance of both algorithms on a new domain -- that of 4-regular graphs, complementing previous analyses done on 3-regular graphs and Erd\H{o}s-R\'{e}nyi random graphs. The obtained dataset will be used to define reasonable criteria when to consider QAOA to have done ``significantly better'', and also to train a machine learning classifier to predict this criterion.

\section{Performances on 4-regular graphs}
\label{QAOAperfs}
In order to systematically compare QAOA against other algorithms, certain choices regarding the QAOA set-up (hyperparameters), the problem test set, and regarding the exact notion of ``better performance'' need to be made.

Regarding the problem test set, as mentioned earlier, we chose to focus on 4-regular graphs, in part because they are relatively easy to generate, yet have not been considered in the literature in the context of comparison to GW.
Although in practice both GW and QAOA would be used by selecting the best performance out of a number of runs, in this work we opted to compare averaged performances as they come with a number of practical and theoretical advantages.
From the theory perspective, both QAOA and GW have been studied theoretically with respect to average performance and this theoretical insight allows us to have a better grasp on  our empirical findings. 
From a pragmatic perspective, due to the fact we can only investigate comparatively small instances, the best cut value over a reasonable number of  $M$ (user-defined) samples would end up in ratios very often equal to optimal performance. Considering
average performances allow us to have a more robust comparison (combined with best performances if relevant).
Note also that in the event that we could explore significantly larger instances, this extreme behavior of seeing mostly optimal results would cease, and observing averages would be more representative. 

Along the same lines, we note that measuring averages, or best-out-of-$M$-shots actually yields, technically speaking, distinct algorithms, with different run-times (due to $M$). Which algorithm is more relevant dependent on the point of view of a practitioner (time to run on a real-hardware, number of function evaluations allowed during optimization...). If the performance measure is best-out-of-$M$-shots, this adds another hyperparameter -- a  degree of freedom -- in the comparison. Technically, our setting is the averaged $M=1$ setting. Using higher $M$ values would involve other aspects of the output distribution (which can also be subject to optimization), yielding a significantly more complicated process.
Finally, we highlight that the main objective of this work is to investigate the capacity of ML methods to detect when QAOA could yield to better performances compared to a classical algorithm. The actual specification of the algorithm is of course important, but still secondary for our study; we note that, from the onset, we had to make a number of modeling selections, including choosing the optimization algorithm, fixing the depth, and in the same vein, choosing to study $M=1$.

Regarding these other important choices in the QAOA algorithm details/hyperparameters, the  depth of the circuit we will allow, and the optimizer used are obviously very relevant. The basis of our optimization is the Nelder-Mead (NM) algorithm, with some QAOA-specific tweaks, detailed shortly. NM is a common choice due to  the relatively small dimensionality of the parameter space. For instance, NM was used in \cite{Guerreschi2019} to give a computational time cost for running QAOA against a classical heuristic. 
Perhaps the most important choice, the allowed circuit depth, is non-trivial. Note that it is known that if the circuit is deep enough, relative to the instance size, we will find true optima (given that the QAOA circuit parameters themselves are chosen optimally). The exact functional scaling of the depth required for this is not known, but as clarified, having the depth depend on the instance size would lead to intractable optimization demands regarding the QAOA parameters -- and defeat the idea of NISQ-friendly constant-depth constructions. However, since we are bounded to relatively small instances due to the cost of quantum simulation, already depths 7-9 may allow us to achieve better performance than what is asymptotically possible for any constant depth\footnote{More precisely, since QAOA finds the exact optimum in the limit of infinite depth, there exists some function $d(n)$, for which QAOA of depth $d(n)$ can achieve performance approximating the optimum arbitrarily well, e.g. better than the GW bound. However, this does not yield an efficient algorithm, even if $d(n)$ grows very slowly, as the finding of the ideal parameters is in principle exponentially expensive in $d(n)$. Nonetheless for the values $n \le 22$ it may be the case that the depth of 9 which we study is larger than $d(22),$ allowing us to nearly-always, or always beat GW in our experiments.  Thus this is a finite-size effect, but of the kind we may care about in practical computing.}. On the other hand, keeping the depth too low may prevent any advantage in the upper range of the graph sizes and test-set sizes that we could tackle. To help us with the choice here, we analyzed the performance of QAOA on a range of depths, as described next. 

Finally, this brings us to the choice of the appropriate definition of ``better performance''.
Relating to previous works, in \cite{Crooks2018b} (although the objective was not to formally define such a criterion), an average improvement of the approximation ratio of 2\% was considered a significant improvement. Our choice specified in the next section is also guided by the data acquisition stage, i.e., the performance analysis on our graph test sets.

\subsection{Simulation methods}

We built a set of randomly generated 4-regular graphs from where we varied the number of nodes $N=11, \ldots 24$, with 20 instances generated per size, i.e., in total 280 instances. 
For each graph, the $C_{max}$ value was found by brute-force so that approximation ratios can be computed. We evaluated both the GW and QAOA algorithm on each instance.
Regarding GW, we ran the standard algorithm and computed the effective performance by dividing the expected cost (estimated using 1000 projections) by the actual $C_{max}$.

Regarding our evaluation of QAOA, more details are warranted. We run QAOA on a quantum simulator, starting from $p=1$ and increasing the depth up to $p=10$. The  algorithm we use  for the QAOA parameter optimization is based on the Nelder-Mead (NM) method, whose performance dramatically depends on the choice of initial values/angles. We embed the NM algorithm  in an optimization procedure outlined in  \cite{qaoaConcentrates} for finding good near-optimal angles for maximizing $ F_p(\vec{\gamma},\vec{\beta})$. The procedure is based on the observation that the evaluation of the QAOA objective function for fixed parameters on graphs generated from a ``reasonable'' distribution concentrates with little fluctuation. This suggests that optimal angles for QAOA from one instance can be used as a starting point for other instances, which is then improved via local search.
Specifically, for a given depth, we find candidates for optimal angles starting from random ``seed'' positions. The newly found angles are  then used as starting points for NM on the next instance. After passing through all graphs in the set, we attempt improving the best achieved ratios for a given $p$. This is done by trying out all angles found for each graph as seeds for further optimisation with NM, but only in the case that the achieved performance was not already better than what  GW achieves. 
This is motivated by the fact that we are predominantly interested in the performance of QAOA assuming the capacity to achieve optimal parameters/angles, and identifying when that performance surpasses GW \footnote{As we explain shortly, we will investigate another, more stringent criterion which requires an objectively higher performance quality, and a larger gap over GW. In principle, we could have forced further optimizations on all instances that failed that criterion (but were already better than GW). We opted not to do so, because such a process would not be feasible in practice. Note that we can guide our optimizations of QAOA relative to the performance of GW (as we can run GW in polynomial time), but we cannot do the same in practice for any criteria which evaluate the actual achieved ratio of QAOA, as this would require computing $C_{max}$.}.

 In those cases where QAOA still underperformed relative to GW, we also run also QAOA with depth 11 and 12, but ultimately, even this did not improve the performance. In the following subsection we present the results obtained using the method above.

\subsection{Results}

We used simulations to first identify the depth of QAOA we wish to focus on. Intuitively we looked for a depth when using QAOA is obviously interesting relative to the performance of GW. Such a decision will  also depend on the choice of the criterion of what makes an algorithm heuristically better than another. In our case, we looked at three figures of merit: minimum and median of the  performance ratio, and overall percentage of instances where QAOA outperformed GW. 
Additionally, we address  the more practically motivated criteria where we aim to identify settings where QAOA does not just outperform GW, but does so with a significant margin, and in the regime where the performance is already much better than the GW lower bounds.
Here we selected the threshold of 98\% for practical and pragmatic reasons: it may be high enough that no poly-time approximation algorithm can do better even on our restricted family of graphs (regardless of the UGC conjecture). This would make it a ``genuinely heuristic'' feature. Yet, it is low enough that we have sufficiently many YES and NO instances for a meaningful analysis.

\begin{figure}[ht]
\begin{center}
\includegraphics[width=0.45\textwidth]{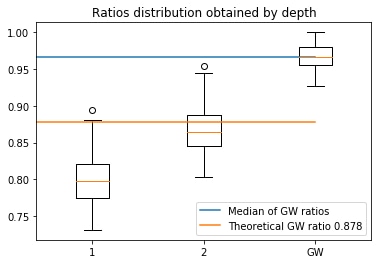}
\includegraphics[width=0.45\textwidth]{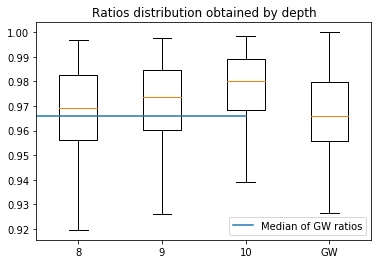}
\caption{Boxplot of ratios for $p=1,2$ and $8,9,10$ as well as the ratio of GW. }
\label{ratio_by_p}
\end{center}
\end{figure}

\begin{table}[t]
\caption{Median and minimal ratios achieved by the QAOA algorithm for a given depth. For comparison we also provide the values achieved by the GW algorithm. We underline the first value where QAOA surpasses GW.}
\label{ratio-table}
\vskip 0.15in
\begin{center}
\begin{small}
\begin{sc}
\begin{tabular}{lcccr}
\br
Depth & Minimal ratio & Median ratio \\
\mr
1    & 0.7312  & 0.7979 \\
2   &  0.8026 &  0.8641 \\
3    & 0.8488 &  0.9050 \\
4    & 0.8777  & 0.9302 \\
5    & 0.8993 &  0.9446 \\
6     &  0.9031 & 0.9523 \\
7     &  0.9209 &  0.9607 \\
8     &  0.9195 &  \underline{0.9692} \\
9     & 0.9259 & 0.9737 \\
10   & \underline{0.9390} &  0.9800 \\
GW & 0.9265 &  0.9658 \\
\br
\end{tabular}
\end{sc}
\end{small}
\end{center}
\vskip -0.1in
\end{table}

Fig.\ref{ratio_by_p} shows boxplots, representing the quartiles of achieved ratios, presented according to depths from 1 to 10. Interesting depths for QAOA start at $p=8$. Indeed, for $p=8,9,10$, QAOA yields a better expected cost than GW on respectively 57.5\%, 67.8\% and 92.14 \% of the instances. Table \ref{ratio-table} displays the obtained minimum and median ratios (averages yield similar conclusions but are generally more sensitive to outliers) on the generated set. At depth 10, we see a clear advantage of QAOA with respect to the minimal ratio criterion. 
\par

\begin{figure}[ht]
\vskip 0.2in
\begin{center}
\includegraphics[width=0.45\textwidth]{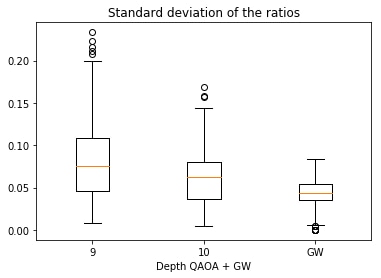}
\includegraphics[width=0.45\textwidth]{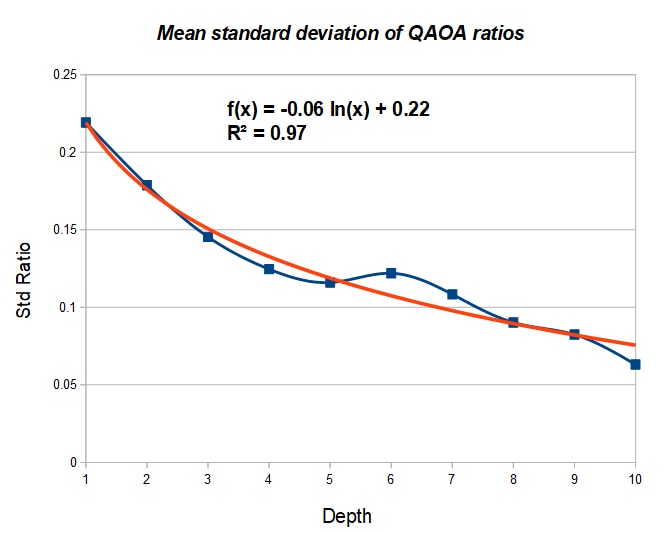}
\caption{Standard deviations of the ratios over the generated graphs obtained by sampling 1000 times for $p=9,10$. The mean standard deviations are respectively: 0.0824, 0.0631 and 0.0432. The second plot is a logarithmic fitting of the mean deviation by depth.}
\label{stds}
\end{center}
\vskip -0.2in
\end{figure}

In addition to the expected cost, we also investigated the standard deviation of the output of the algorithms. 
This quantity is particularly relevant when randomized algorithms are used as subroutines in other algorithms (e.g. as steps in local search), in which cases more stable performance (with a worse mean) may be preferred over on-average-better, but much less reliable performance. 
We took the optimal angles obtained at depths 9 and 10, from which we sample the circuit 1000 times and compute the MaxCut cost. For GW, we use the 1000 random projections obtained for computing the expected ratios. From Fig.\ref{stds}, we observe that the QAOA output is more spread out around the average than GW, and would require higher circuit depth to decrease it. 
In \cite{QAOA} it was shown that the upper bound on the spread (specifically, the variance) of the output distribution  is dictated by a term of the form $(v-1)^{2p+2}$, where $v$ is the constant degree of the graph (and $p$ is the depth). Due to this form, we applied a logarithmic fitting to the found variances, to estimate at which depths QAOA spread is guaranteed to concentrate more than GW. Numerically, we estimate this to occur after depth 20.
However going to these higher depths did not yield better expected ratios (and did make optimisation significantly more costly and unstable),  
so we restricted the depths to $p \leq 10$ in the subsequent steps of our study.
\par
Last but not least, we are interested in defining when QAOA is a good heuristic.  In \cite{Crooks2018b}, the average approximation ratio (given on graph instances of similar size) at depth 8 was between 0.98 and 0.96, with an improvement close to 2\% against GW. We then choose arbitrarily to define QAOA as a significanly good heuristic when QAOA yields a ratio higher than 0.98 and an improvement against GW by at least 2\%. This occured on 24 instances (8.5\%) and gives rise to potentially good situations when QAOA is more suited to be advantageous. Indeed, those will be the cases where QAOA is nearing perfect performances. At most the returned cut values will differ by 1 from $C_{max}$ (on small instances) on average. Although this criterion may be pretty strict, it is interesting for algorithm selection. Having defined labels for our generated set of instances, we inferred when QAOA is more suited for applications.

\section{Characterizing QAOA advantage}
\label{AlgoSel}

As indicated in the previous section, we identify two basic regimes of interest. The first criterion (1) -- the \emph{GW vs. QAOA} criterion,  is the basic question: can we identify those instances where the ratio achieved by GW is surpassed by that of QAOA at all. 
The second is a question whether QAOA performs remarkably well in a more general sense -- whether it is a ''good heuristic'', so to say. 
Note that what values an algorithm has to attain to be considered a ''good heuristic'' is of course arbitrary, and our decision of setting this threshold at 0.98 is motivated by three main factors: this is the highest value where we obtain non-trivial sample sizes where QAOA also outperforms GW by a substantial margin (2\%); this is also likely in the NP-hard regime (if the performance could be attained for \emph{all} graphs), even when restriced to 4-regular graphs. Finally, achieving objectively high performance values intuitively reduces the chances that other heuristic algorithms will generically be better at those instances.

Thus our second criterion (2) -- \emph{the good heuristic criterion} is: does QAOA attain performance above $0.98$ with an advantage margin of at least $0.02$ over GW?

Note that deciding the first criterion is trivial if we are allowed to actually run QAOA. The second criterion may be more problematic since it implies we can guess whether QAOA will do better than $0.98$.

Since we are interested in developing a methodology which helps us decide whether to run QAOA at all, we are only interested in detection methods which can be run efficiently on a classical device. So one may wonder whether characterizing instances with respect to the two criteria can be done algorithmically in polynomial time on a classical computer.
\par

Regarding the two criteria, although we do not make any hard claims specifically for our choices, we point out that it is not difficult to see that finding exact predictors of performance is in general exceptionally difficult. For instance, deciding if GW does better than some threshold $\alpha < \beta$ (where $\alpha$ is the GW bound) is already NP-hard under the UGC\footnote{Under the unique games conjecture (UGC), the NP-hard bound for approximation coincides with $\alpha$ achieved by the GW algorithm \cite{Khot:2002:PUG:509907.510017,Khot:2007:OIR:1328722.1328735}.} .
Given access to an algorithm deciding this, and to the cut value $r$ returned by the GW algorithm, in the case the output is YES, we have a cut value above $\beta C_{max}$, where $C_{max}$ is the true optimum. In the NO case, we can conclude that there exists a cut with value $ (r/\beta) \geq (\alpha/\beta) C_{max}> \alpha C_{max}$ (since $\alpha C_{max} < r < \beta C_{max}$). This results in an overall $ \min(\beta, \alpha/\beta)$-approximation algorithm, beating GW\footnote{Note, already estimating the cut size (rather than outputting the cut itself) better than the GW bound is NP-hard, under the UGC \cite{Khot:2002:PUG:509907.510017,Khot:2007:OIR:1328722.1328735}.}. Similar arguments can be applied to other algorithms and criteria as well. For instance, assuming QAOA is an $\alpha'$-approximation algorithm (we know $\alpha'~\leq 5/6+\delta$ for constant $p$ and any $\delta >0$) deciding if QAOA does better than the $GW$ bound $\alpha+\epsilon$ by any margin $\epsilon$ is likely hard. YES results give an $\alpha+\epsilon$ approximation, whereas NO yields an $\alpha'/\alpha$ approximation, which beats GW if $\alpha' > \alpha^2 \approx 0.77$. This is significantly lower than the plausible bound of $5/6$ \cite{variationalObstacles}. Thus, unless QAOA is weaker than a $0.77-$approximation algorithm for all constant depths, deciding whether it beats the GW bound on a given instance would imply $BQP \subseteq NP,$ under the UGC.

In the more general case when we compare QAOA against more complicated heuristics, for which bounds may be unknown, proving formal claims may be even more difficult, and as we explain shortly, arguably less useful.

\subsection{Machine learning for performance prediction}
The reasons above are an important factor why we resort to a machine learning (ML) approach, specifically supervised learning, to decide our criteria (1) and (2). Another one is because machine learning methods are also more robust, and flexible. We note that exactly the same method we employ here for characterising the performance of the idealized QAOA relative to GW can be used with any other classical heuristic, and with real-world, noisy QAOA implementations.

Our method is inspired by the approach  in \cite{DunningEtAl2015} where the authors inferred a ranking between different classical MaxCut heuristics, gathering instances from many MaxCut and QUBO libraries. However, due to the needs of quantum simulation, we could only provide more modest  sample set and instance sizes. 

In essence, we prepare a dataset of instances for which we compute the criterion value (NO/YES, or  0/1 for each criterion) -- in supervised learning this is called a \emph{label}. We then train a ML model to fit this value, using a subset of data -- the training set; the performance reported is obtained by applying the model on the testing set. Using 4-fold cross-validation in our case gives an idea of generalisation and robustness of the model.

When relying on machine learning, a key step is the identification of the \emph{features}, that is, the pieces of information (i.e. properties of the instance) based on which we will be training the model and making a decision. While the graph itself is a feature implicitly containing all the information we may care about, it is almost always useful to pre-compute many other derived properties (e.g. the graph density, or some property of the spectrum of the graph adjacency matrix) and use them directly in training the ML classifier. Note that in principle the performance itself could be a feature, however, we limit ourselves only to use features which can be computed classically, and efficiently relative to the running of the QAOA algorithm -- the entire idea is to decide \emph{whether} to employ the quantum device at all.

In general, our approach incorporates the 3 standard phases when employing ML: preparing the dataset (incl. pre-computation of the features, the running of the QAOA to compute the true criterion value, etc.), the training of the ML model, and evaluation.
This process was iterated a number of times before the most informative features were identified. In our analysis we also discuss which features contributed the most to prediction accuracy; this may also be informative for theoretical analyses of QAOA performance.

\noindent \textbf{The learning model -} In our study we have initially considered a number of models to perform prediction. We have experimented with traditional ML models such as tree-based  gradient boosting, random forests, LightGBM and XGBoost \cite{rf,xgboost, lightgbm} -- which are simple and would have high interpretability (meaning we can infer something about why the model makes certain decisions). However, accuracies were very low (less than $63\%$). This presumes that a more complicated ML machinery is needed, especially when dealing with small datasets.

As we are predominantly interested in high accuracy for these computationally hard predicates, vital for a practical advantage, we opted to deal with relatively complex models, the hyper-parameters of which were optimised using automated (auto-ML) techniques. Specifically, we built our models using the TPOT library \cite{OlsonGECCO2016}, which builds the models using an evolutionary strategy. A clear advantage of such auto-ML methods is that they offer a significantly enhanced level of flexibility for the user, as much of the vital hyper-parameter optimization is taken care of automatically. This suggests that, with very little interventions from the user, the same techniques we provide could be used for noisy QAOA, or comparisons against very different types of heuristics. Such flexibility is a key desired feature of automated algorithm selection.

\noindent \textbf{Features utilized for prediction -} In our analysis we do not employ the raw graph description as a feature as it is carries a lot of irrelevant information (e.g. both algorithms are invariant under graph isomorphisms). 
We used some of the features proposed in \cite{DunningEtAl2015} suited for unweighted graphs, while generating others for our results via a trial and error process. 
In the final analysis, we investigated the potential 20 features, listed in appendix \ref{feat}. These are grouped in three classes: (i) graph spectral properties, (ii) subgraph characteristics (e.g., maximum independent set size), and (iii) certain GW performance features (e.g., normalized value of the relaxed problem). This third class of features is of course GW-specific, but it should not be surprising that the performance of GW itself carries a lot of information about whether QAOA will do better. 
We note that some of the features in group (ii) are actually NP-hard to compute, so we cannot expect their exact values to be used in practice to guide our decisions. While we nonetheless investigated their value as predictors (we show they are absolutely unnecessary), the best results we report only utilize efficiently computable features. Criterion (1) is tackled before (2).

\subsection{Predicting Criterion (1): QAOA vs. GW}

We computed the features for each of our 280 graph instances, and set the label to 1 when GW outperforms QAOA (at depth 10) in terms of the approximation ratio, and 0 otherwise. Note that to compute this particular label, the computation of the  actual optimum is not necessary, but the running of QAOA is.

We then built a classifier by 4-fold stratified cross-validation and optimizing the average balanced accuracy over the validation sets;
the data was split into 4 subsets, and the model is trained on 3 of them while being tested on a fourth. Stratified sampling is used in order to maintain the ratios of the labels in the sets. This procedure is performed 4 times and performances are averaged.  k-fold cross-validation is a commonly used method to build ML models on small datasets while still giving a robust idea of performance and generalisation features.

Since our data is highly biased (QAOA outperforms GW significantly, more than half of the time), the relevant measure of performance is the so-called balanced accuracy \cite{balancedacc}, which uniformly averages the performances for YES and NO instances.

In our analysis, we initially used all 20 features, and then proceeded to prune out a smaller number with the highest impact on accuracy.
In the end, we have identified two features which alone enable the same balanced accuracy, as all features combined, and these are:
\begin{itemize}
\item expected\_costGW\_over\_sdp\_cost: the expected cost over the 1000 random projections per instance divided by the cost of the relaxation $C_{rlx}$,
\item std\_costGW\_over\_sdp\_cost: the standard deviation of the cost approximated using 1000 random projections divided by $C_{rlx}$.
\end{itemize}
In other words, to achieve the +96\% accuracy we report, we only need to use these two features. The model is described in appendix \ref{crit1}.

A few comments are in order; first, note that both of these features exclusively characterise the output of GW -- this is not too surprising as the criterion ``better than GW'' strongly depends on GW performance.
For instance, the first of these two features is likely closely correlated to the actual performance of GW in terms of the approximation ratio (it is the exact value when $C_{rlx} = C_{max}$).
However, we can also understand the performance analysis of GW as a feature extraction mechanism, which identifies the properties of graphs which also make them suitable for QAOA. 
This is indeed the case, as our classification algorithm correctly predicts QAOA advantage even in many instances where GW did exceptionally well, and also, correctly predicts that QAOA underperformed also when GW did ``badly'' as well -- in other words, we capture more than the quality of  performance of GW.
Also importantly, this shows that NP-hard features are really unnecessary to decide criterion (1).

\begin{figure}[ht]
\vskip 0.2in
\begin{center}
\centerline{\includegraphics[width=\textwidth]{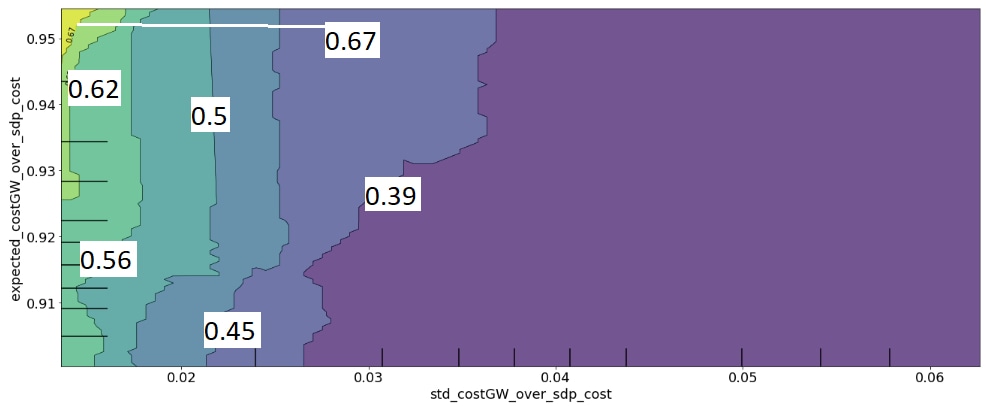}}
\caption{Partial dependence plot for the classifier built on two features. The contour lines from left to right represent probabilities thresholds : 0.67, 0.62, 0.56, 0.5, 0.45, 0.39.}
\label{pdp}
\end{center}
\vskip -0.2in
\end{figure}

We analyzed these two features and their relationship in more detail.
By looking at the partial dependence plot of the classifier in Fig.\ref{pdp}, we quantified how the predicted probabilities evolve according to those two features. 
In more details, our classification model outputs probabilities of assignments, and not the labels themselves. The closer the probability is to 1, the more likely the output label will be a 1.
From the plot, we observe that the smaller the standard deviation of the cost applied on the random projections, and the closer the expected cost to $C_{rlx}$ is, the higher is the probability that GW outperforms QAOA. But this also means that the GW performance is not far from an actual optimum since $C_{rlx}$ upper bounds $C_{max}.$
In order to understand the importance/impact of these features, we also performed a so-called permutation importance analysis, in essence based on permuting the values of a column in the dataset corresponding to one feature, and measuring the impact on the overall classifier performance.
When this measure was applied to the two features above, we noted that performances decrease by $0.4283 \pm 0.1843$ for the standard deviation feature and $0.0533 \pm 0.0743$ for the expected cost, indicating the prior is more informative. 
\par

While showing that GW performance influences whether QAOA outperforms it is unsurprising, it does open the obvious question of whether it is possible to find other sufficiently informative features. We would especially be interested in finding features to be used in place of the standard deviation feature, which requires the costly random projection evaluations.

To this end, we have rerun the analysis from scratch, running through a new cycle of automated hyper-parameter optimization, trying to match the performance without using these two GW-specific features  -- including using the NP-hard features. However, we have failed to achieve performance close to when those features were available. To understand this further, we have applied a regression approach, trying to predict the values of the expensive feature std\_costGW\_over\_sdp\_cost  from the other features. Note that if this were possible, then the other features indeed did contain sufficient information, even without these expensive features being explicitly present. We analyzed the explained variance of the constructed regressor, and obtained a low value of 0.7693. With this result, it became apparent that we should not expect to be able to recover the same levels of accuracy without these costly (but still efficiently computable) features, using our machine learning model.
In principle, of course all features should be distillable from raw graph data, and in future work it will be valuable to achieve as good or better performance relying on better tuned models, and cheaper features.

To summarize the results of this section, we can report that the prediction performance for criterion (1), which can be achieved when many NP-hard graph-theoretical features are available, along with the less costly features, can easily be matched using just properties derived from the tractable GW performance.
It is worth to note that the generation of the datasets for this prediction is also tractable (albeit, expensive), as it requires only the running of QAOA and of GW.

\subsection{Predicting Criterion (2): QAOA as a High-Performing Heuristic}

For this analysis, we considered the same dataset but the label was changed. We labeled with 1 the instances were the QAOA ratio exceeded 0.98 and where the gain was at least 2\% over the performance of GW. Note that to generate this label, we did need the true $C_{max}$ values which we computed by brute-force. We will discuss this issue presently.
We built a classifier which utilized all the features we discussed previously, including the features relying on the random projections and the NP-hard graph-theoretical features, which lead to 0.8255 in balanced accuracy. The analysis was done again using 4-fold cross validation.
This performance provides a benchmark which we would ideally like to achieve using only those features which can be computed in polynomial time.

\begin{figure}[ht]
\vskip 0.2in
\begin{center}
\includegraphics[width=0.7\textwidth]{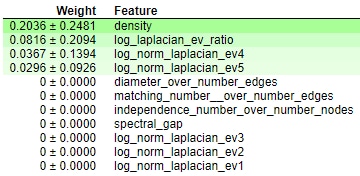}
\includegraphics[width=\textwidth]{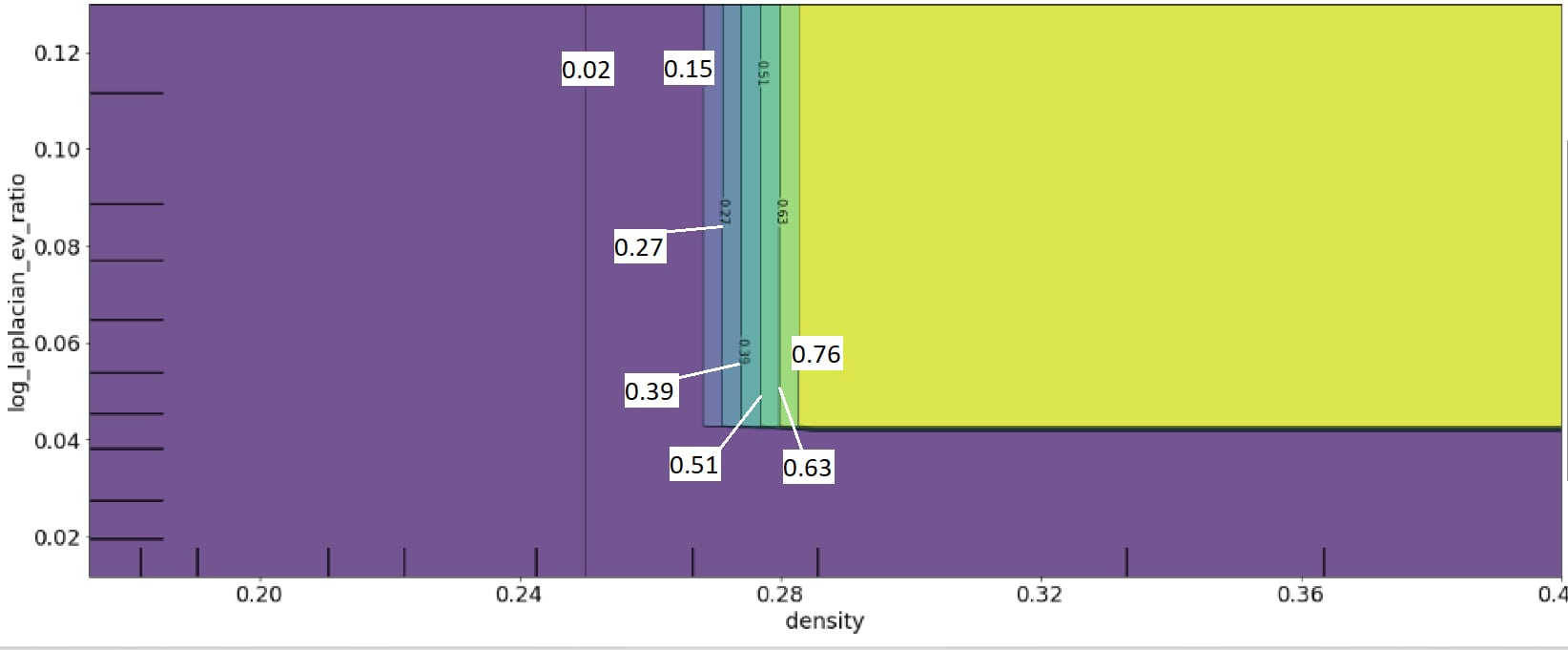}
\caption{Partial dependence plot for the discriminator of significant gain with QAOA with its permutation importance.
The yellow box correspond to probabilities higher than 76\%, so likely to be labeled 1 or being in the second regime.
 The contour lines from right to left represent probability thresholds as follows: 0.76, 0.63, 0.51, 0.39, 0.27, 0.15, 0.02.
}
\label{pdp2}
\end{center}
\vskip -0.2in
\end{figure}

To do so, we performed importance analyses of the features and built a classifier (described in appendix \ref{crit2}) using a subset of the features, first disregarding the GW-dependent features.
With this collection of features, the obtained accuracy was essentially as good: 0.8236 and with a recall (intuitively the ability of the classifier to find all the class 1 instances) of 0.7917. 
We analyzed the feature importance shown in Fig.\ref{pdp2}, which shows that not only the GW-specific features can be dropped, but also that the NP-hard graph-theoretic features are unnecessary. 
So we can discard features (ii) and (iii) in the second regime, keeping the green-highlighted ones -- and these are all computationally efficiently computable features, yielding an efficient classifier. For completeness, we did train and test the classifier using only these features, resulting in the accuracy of 0.8236, as expected.

Further, we note that the reported performance is very good from a ML/classification point of perspective (in an absence of other well-established benchmarks).

In more detail, we report that the density and the ratio between the largest eigenvalues of the adjacency matrix of the graph seem to be  most important for deciding criterion (2), while it seems that demanding the gap of 2\% between GW and QAOA significantly diminished the explicit dependence on the GW features, which was so prominent for criterion (1). 
To further investigate this, we constructed a partial dependence plot, which revealed that regular graphs with more nodes are overrepresented in the high-performance regime. Further, those cases correspond to the case when the largest eigenvalue of the Laplacian matrix is at least 1.05\% times the second largest eigenvalue. Finally, we note that problem density was also found to matter for QAOA performances in satisfiability problems \cite{reachability}, defined as the ratio between the number of clauses and the number of variables. A similar type of property (or feature, in machine learning terminology) also stood out in the analysis of the performance of our algorithm as important, in particular for the second criterion. This is suggesting that density-type features could be important for algorithm selection, which involves QAOA, more generally.

In summary, from our 82\% accurate ML model, we found interesting graph properties which influence whether a QAOA approach will result in a high-performing heuristic. This is mainly dependent on the Laplacian spectrum and the density. Using ML explainability, we quantified best significantly handled graph instances. Specifically, the density and the ratio between the largest eigenvalues of the adjacency matrix of the graph were the most informative features in this regime.
This highlights a possible direction for further analytic and experimental investigations of QAOA performances. 

\section{Discussion}
\label{discuss}

In this work we raise the problem of algorithm selection when quantum approaches, specifically the QAOA algorithm, are considered. 
We were interested in detecting instances where quantum advantage over a classical algorithm is obtainable, as characterised by two regimes: when QAOA simply outperforms the classical algorithm, in this case the Goemans-Williamson algorithm run on the MaxCut problem (the first regime); and a practically more significant regime, where the quantum heuristic yields a ``very-high-quality'' output, while significantly beating the classical algorithm (second regime). Our choices of the criteria were guided by practical and pragmatic considerations, based on our analysis of a dataset of graphs. 

The algorithm selection problem boils down to the design of a classification algorithm, which can efficiently detect whether a given graph satisfies the first, and/or the second criterion. To this end, we developed a machine learning approach for these types of tasks. Our choice to resort to machine learning  was motivated by a number of considerations; first, machine learning methods are flexible, and offer confidence that our approach can equally be applied when using QAOA (or other quantum algorithms, for that matter) in other regimes (e.g., when considering real-world noise), when compared other classical algorithms. Second, we provided simple complexity-theoretical arguments why exact decisions are likely computationally intractable. Third, using ML methods allowed us to employ explainability techniques to identify features of graphs which make them better suited for a quantum treatment, which may guide other more theoretical research.

Finally, to our knowledge, the use of ML techniques to analyze the performance of quantum methods has only previously been employed in \cite{qwalkadv} (albeit for a different purpose); yet, we believe ML will play an ever increasing role in helping us identify interesting quantum heuristics, so we hope our work may motivate more studies in this direction. For instance, machine learning methods of the type we propose could conceivably be used to identify and characterise what types of datasets are best suitable for quantum-variational-circuit-based classification, which is arguably one of the key question in NISQ-oriented quantum-enhanced machine learning domain.  
 
\par
In the process of analysing  QAOA and GW performance, we found further evidence that QAOA can provide genuine advantages over the GW algorithm-- specifically, we have shown that that at depth $p=10$, QAOA outperforms GW on most instances of 4-regular graphs up to size 24. 
With respect to predicting advantages, we constructed a model yielding an accuracy of  96\%  for the first regime -- the key features here depend on the output of GW, and other features were less significant. For the second regime, we achieved a model with 82\% balanced accuracy, and further used explainability methods to elucidate which graph features influence QAOA performance the most. In this regime, spectral properties of the graph, and basic graph density were most influential.
We note that already the accuracy of 82\% (4-fold cross-validated) would yield substantial optimizations in the use of quantum resources in any larger-scale optimization effort, of the type we may expect in, e.g., industrial applications.

\par
We see numerous possibilities for future work. First, it would be interesting to perform a similar analysis on different graph families and identify when to use QAOA. 
Second, we could consider using Graph Sparsification \cite{sparsity}, that converts a weighted graph into a sparser one preserving all cuts up to multiplicative error. This can be done classically in near-linear time, and quantumly even in sublinear time. We would then study if performances of both GW and QAOA are improved on the sparser instance.
Third, an obvious limitation to this type of study is the size of the graph we can handle; to this end it would be interesting if divide-and-conquer type methods explored in \cite{qdivide} can be utilized to increase the size of datasets we can consider.
Additionally, in our work, we did not tackle the important question of how the choice of QAOA optimization procedures influences the advantage gained. It is also important to consider how realistic, or real, noise effects affect performance. Noise and device errors will of course impact the performance of QAOA \cite{analysisNoise1, analysisNoise2, qaoasycamore}, but all this will improve through modern error mitigation techniques \cite{Zhang2020} and even more so with the increase of device fidelities. As long the devices prevent good performance, classical algorithms will in general beat the quantum approaches. Nonetheless, since we employ quite general machine learning methods, the algorithm selection methodology remains the same. Indeed, in general, each improvement of hardware constitutes an effectively new optimization algorithm with different learning features which may beat the classical system in different instances. Since we use ML methods, nothing in our process actually relies on actual ideal implementations of QAOA, and exactly the same algorithm could be used in all noisy settings. Plus, it is also likely that the performance of our algorithm would be at least as good as in the ideal case (in the noisy cases where the quantum system is seldom better, a larger bias to classical algorithms direclty improves performance of the algorithm selection process). We could even go further by considering a fleet of hardware of different types and characteristics. Adding hardware-related features to decide whether to run QAOA or not on a specific machine
for solving an instance could be then studied in future work. 

Finally, more recent contributions also provide new modifications of QAOA to make it more suitable for standard real-world optimization objectives \cite{gibbsQAOA,cvarQAOA}. We believe our methods can equally be applied in those settings.

\section*{Acknowledgements}

This work was supported by the Dutch Research Council(NWO/OCW), as part of the Quantum Software Consortium programme (project number 024.003.037).
CM and VD acknowledge support from Total in providing access to the Atos Quantum Learning Machine for their simulations. VD and CM are grateful to Thomas B\"{a}ck, as well as anonymous reviewers for a careful reading of the manuscript, and for useful suggestions, and to Ronald de Wolf for useful comments and suggestions for the manuscript, for pointing out new references, and for discussions regarding the hardness of prediction of performance of approximation algorithms. 

\section*{References}
\bibliography{references}

\appendix
\section{Simulations}

\textbf{Generated graphs}
The following code reproduces the instances:
\begin{lstlisting}[language=Python]
from networkx import random_regular_graph
gs = []
for n in range(11,25):
 for i in range(10):
  gs.append(random_regular_graph(4,n,i*10))
  gs.append(random_regular_graph(4,n,(i+1)*11))
\end{lstlisting}

\section{Features}
\label{feat}

\subsection{Common features related to regular graphs and the spectrum of the Laplacian matrix}
\label{feat1}
\begin{itemize}
\item density: percentage of the number of edges in a complete graph with the same number of vertices. For regular graphs, this is strongly correlated to the number of vertices.
\item log\_norm\_laplacian\_ev1: logarithm of  largest eigenvalue of the laplacian normalized by the degree.
\item log\_norm\_laplacian\_ev2:  logarithm of the second largest eigenvalue of the laplacian normalized by the degree.
\item the same for the third, fourth and fifth largest eigenvalues.
\item log\_laplacian\_ev\_ratio:  logarithm of the ratio between the two largest eigenvalues.
\item spectral\_gap: the second smallest eigenvalue of the laplacian.
\end{itemize}

\subsection{(ii) Set numbers for graphs}
\label{feat2}

\begin{itemize}
\item independence\_number\_over\_number\_edges: cardinality of a largest independent set of nodes in the graph normalized by the number of edges.
\item matching\_number\_over\_number\_edges: cardinality of a maximum matching (size of a maximum independent edge set) normalized by the number of edges.
\item diameter\_over\_number\_edges: diameter is the maximum eccentricity normalized by the number of edges.
\item domination\_number\_over\_number\_nodes: number of vertices in a smallest dominating set divided by number of nodes.
\item zero\_forcing\_number\_over\_number\_nodes: minimum cardinality of a zero forcing set divided by number of nodes.
\item power\_domination\_over\_number\_edges:  minimum cardinality of a power dominating set divided by number of nodes.
\end{itemize}

\subsection{(iii) Features related to the relaxed solution of the semi-definite program in GW and the random projection routine}
\label{feat3}

\begin{itemize}
\item percent\_cut: ratio between $C_{rlx}$ and the number of edges.
\item percent\_positive\_lower\_part\_relaxation\_solution: percentage of elements that are positive in the relaxed solution after Cholesky factorization.
\item percent\_close1\_lower\_part\_relaxation\_solution:  percentage of elements that are less than .1 in absolute value in the relaxed solution after Cholesky factorization.
\item percent\_close3\_lower\_part\_relaxation\_solution: percentage of elements that are less than .001 in absolute value in the relaxed solution after Cholesky factorization.

\item expected\_costGW\_over\_sdp\_cost: the expected cost over the 1000 random projections per instance divided by the cost of the relaxation $C_{rlx}$,
\item std\_costGW\_over\_sdp\_cost: the standard deviation of the cost approximated using 1000 random projections divided by $C_{rlx}$.
\end{itemize}

\section{Models}
\label{mlmodels}

We give below the code of the pipeline constructed by TPOT for the two Criteria. Each algorithm name (either data transformer or classifier) can be found in the Scikit-learn library \cite{scikit-learn}. Names ending in NB are Naive-Bayes classifiers.

\subsection{Criterion (1)}
\label{crit1}
\begin{lstlisting}[language=Python]
exported_pipeline = make_pipeline(
    make_union(
        Normalizer(norm="l2"),
        FunctionTransformer(copy)
    ),
StackingEstimator(estimator=\
KNeighborsClassifier(n_neighbors=41,\
 p=1, weights="uniform")),
MultinomialNB(alpha=0.1, fit_prior=False)
)
\end{lstlisting}

\subsection{Criterion (2)}
\label{crit2}

\begin{lstlisting}[language=Python]
exported_pipeline = make_pipeline(
SelectPercentile(score_func=f_classif, \
percentile=95),
StackingEstimator(estimator=\
DecisionTreeClassifier(criterion="entropy", \
max_depth=2, min_samples_leaf=13, \
min_samples_split=9)),
Binarizer(threshold=0.25),
StackingEstimator(estimator=\
KNeighborsClassifier(n_neighbors=8, p=1,\
weights="uniform")),
StackingEstimator(estimator=\
BernoulliNB(alpha=10.0, fit_prior=False)),
StackingEstimator(estimator=GaussianNB()),
MultinomialNB(alpha=0.001, fit_prior=False)
)
\end{lstlisting}

\end{document}